\input harvmac
\def \pa{\partial}

\def\mpla{{Mod.\ Phys.\ Lett.\ }{\bf A}}
\def\npb{{Nucl.\ Phys.\ }{\bf B}}

\def\plb{{Phys.\ Lett.\ }{\bf B}}
\def\prd{{Phys.\ Rev.\ }{\bf D}}
\def\prl{Phys.\ Rev.\ Lett.\ }

\def\bbeta{\bar\beta}
\def\tbeta{\tilde\beta}
\def\Phidot{\dot\Phi}
\def\Phiddot{\ddot\Phi}
\def\Mdot{\dot M}
\def\Mddot{\ddot M}
\def\Mrdot{\dot M_r}
\def\Mrddot{\ddot M_r}
\def\Mrt{M_r^{(3)}}
\def\Mt{M^{(3)}}
\def\Gdot{\dot G}
\def\Bdot{\dot B}
\def\Gdotdot{\ddot G}
\def\Bdotdot{\ddot B}
\def\gdot{\dot g}
\def\e{\epsilon}
\def\a{\alpha}
\def\b{\beta}
\def\g{\gamma}
\def\d{\delta}
\def\r{\rho}
\def\k{\kappa}
\def\l{\lambda}
\def\s{\sigma}
\def\t{\tau}
\def\m{\mu}
\def\n{\nu}
\def\o{\omega}
\def\tW{\tilde W}
\def\tX{\tilde X}
\def\del{\nabla}
\def\pa{\partial}
\def\frak#1#2{{\textstyle{{#1}\over{#2}}}}
{\nopagenumbers 
\line{\hfil LTH467}  
\line{\hfil hep-th/9911064}
\line{\hfil Revised Version}
\vskip .5in
\centerline{\titlefont $O(d,d)$ invariance at two and three loops}

\vskip 1in
\centerline{\bf I.~Jack and S.~Parsons}
\medskip
\centerline{\it Dept. of Mathematical Sciences,
University of Liverpool, Liverpool L69 3BX, UK}
\vskip .3in
We show that in a two-dimensional $\sigma$-model whose fields only depend on
one target space co-ordinate, the $O(d,d)$ invariance of the conformal 
invariance conditions observed at one loop is
preserved at two loops (in the general case with torsion) and at three loops
(in the case without torsion). 
\Date{Nov. 1999}}
\newsec{Introduction}
Duality invariance has proved to be an immensely fruitful concept in string 
theory. A particular aspect of duality which has been valuable in string 
cosmology (for a recent comprehensive review of this subject, with an extensive 
list of references, see Ref.~\ref\eddie{J.E.~Lidsey, D.~Wands and E.J.~Copeland,
hep-th/9909061}) is the concept of $O(d,d)$ invariance. This is displayed 
in the case where the fields 
depend only on time in an arbitrary number of dimensions (time and $d$ spatial
dimensions). It was discovered some time ago\ref\mva{K.A.~Meissner and 
G.~Veneziano, \plb267 (1991) 33; \mpla6 (1991) 3397}  that the lowest order
string effective action (in the simplest case with only
the metric, dilaton and antisymmetric tensor field) then exhibits continuous, 
global $O(d,d)$ invariance. (This is also reminiscent of the $O(d,d)$
invariance previously observed in the context of string compactifications\ref
\giva{A.~Giveon, E.~Rabinovici and G.~Veneziano, \npb322 (1989) 167\semi
A.~Shapere and F.~Wilcek, \npb320 (1989) 669\semi
A.~Giveon, N.~Malkin and E.~Rabinovici, \plb220 (1989) 551}.)
The $O(d,d)$ invariance of the string action was later found to persist in the 
presence of matter or gauge fields\ref\gasp{M.~Gasperini and G.~Veneziano,
\plb277 (1992) 256\semi
S.F.~Hassan and A.~Sen, \npb375 (1992) 103}.
Now the conformal invariance conditions for 
the fields are related to the field equations of the string action, and so, at
least at lowest order, the $O(d,d)$ invariance can be used to transform
between different conformal backgrounds.
Duality invariance in general can be 
understood as a consequence of an isometry in an underlying theory\ref\busch{
T.~Buscher, \plb194 (1987) 59}; and the 
$O(d,d)$ invariance can be viewed as the result of gauging $d$ abelian 
isometries\ref\giv{A.~Giveon and M.~Ro\v cek, \npb380 (1992) 128}. 
In Refs.~\mva\ and \ref\sen{A.~Sen, \plb271 (1991) 295}\
it was argued that the $O(d,d)$ invariance should be
maintained to all orders. 
In Ref.~\ref\meis{K.A.~Meissner, \plb392 (1997) 298}\
the two-loop string action was considered in the context of 
fields depending only on time, and it was shown that after 
a suitable field redefinition it could be written in an explicitly 
$O(d,d)$ invariant form. However, it was pointed out in 
Refs.~\ref\haag{P.E.~Haagensen and K.~Olsen,
\npb504 (1997) 326\semi
P.E.~Haagensen, hep-th/9708110}\ (in the 
context of $T$ duality) that the 
invariance of the action does not 
manifestly guarantee that the conformal invariance conditions
transform appropriately.
Therefore it seems to us that it 
is necessary 
to check explicitly the transformation properties of the 
conformal invariance conditions in order to 
complete the proof that $O(d,d)$ invariance is preserved. At one loop this has 
been done in Ref.~\mva. 
The main purpose of this paper is to verify the invariance 
at two loops (with torsion) and at three loops (without torsion).
We shall find various subtleties 
which do not arise in checking the invariance of the action.

\newsec{The one-loop case}
The two-dimensional non-linear $\s$-model is defined by the action
\eqn\sigact{
S={1\over{2\pi\alpha'}}
\int d^2x\left\{\sqrt\g g_{\m\n}\pa_{\a}x^{\m}\pa^{\a}x^{\n}
+\e^{\a\b}b_{\m\n}\pa_{\a}x^{\m}\pa_{\b}x^{\n}-2\alpha' \sqrt\g 
R^{(2)}\phi\right\},}
where $\alpha'$ is the usual string coupling, which we shall henceforth
set to 1 for convenience,
$g_{\m\n}$ is the metric, 
$b_{\m\n}$ is the antisymmetric tensor field 
and $\phi$ is the dilaton. $\g_{\a\b}$ is the metric on the two-dimensional
worldsheet, $\g=\det\g_{\a\b}$, $\e^{\a\b}$ is the two-dimensional alternating
symbol and $R^{(2)}$ is the worldsheet Ricci scalar.
Conformal invariance for the $\s$-model requires the vanishing of the
three functions $\bbeta^g$, $\bbeta^b$ and $\bbeta^{\phi}$, which are defined 
by\ref\ark{
A.A.~Tseytlin, \plb178 (1986) 34\semi
G.M.~Shore, \npb286 (1987) 349\semi
G.~Curci and G.~Paffuti, \npb286 (1987) 399}
\eqn\Ea{\eqalign{
\bbeta^g_{\m\n}=&\beta^g_{\m\n}+2\del_{\m}\partial_{\n}\phi
+2\del_{(\m}S_{\n)},\cr
\bbeta^b_{\m\n}=&\beta^b_{\m\n}+H^{\r}{}_{\m\n}\partial_{\r}\phi+
H^{\r}{}_{\m\n}S_{\r},\cr
\bbeta^{\phi}=&\beta^{\phi}+(\partial\phi)^2+\partial^{\r}\phi S_{\r}.\cr}}
Here $\beta^g$, $\beta^b$ and $\beta^{\phi}$ are the renormalisation group 
$\beta$-functions for the $\s$-model, and $H_{\m\n\r}$ is the field 
strength for $b_{\m\n}$, defined by $H_{\m\n\r}=3\del_{[\m}b_{\n\r]}$.
The vector $S^{\m}$ arises in the process of defining the trace of the 
energy-momentum tensor as a finite composite operator, and can be computed
perturbatively. At one loop we have
\eqn\Eb{\eqalign{
\beta^{g(1)}_{\m\n}=&R_{\m\n}-{1\over4}H_{\m\r\s}H_{\n}{}^{\r\s},\cr
\beta^{b(1)}_{\m\n}=&-{1\over2}\del^{\r}H_{\r\m\n},\cr  
\beta^{\phi(1)}=&-{1\over2}\del^2\phi-{1\over{24}}H^2,\cr}}
where $H^2=H_{\k\l\r}H^{\k\l\r}$, and $S^{\m}=0$. 
The conformal invariance conditions can be derived at this order from the
string effective action 
\eqn\Ec{
\Gamma^{(1)}=\int dx^{d+1}\sqrt{-g}e^{-2\phi}\left\{R-\frak{1}{12}H^2
+4(\partial\phi)^2\right\}.}
To be more precise, we have
\eqn\Ed{\eqalign{
{\delta\Gamma^{(1)}\over{\delta g^{\m\n}}}
=&\bbeta^{g(1)}_{\m\n}+g_{\m\n}\tbeta^{\phi},\cr
{\delta\Gamma^{(1)}\over{\delta b^{\m\n}}}=&-\bbeta^{b(1)}_{\m\n},\cr
{\delta\Gamma^{(1)}\over{\delta \phi}}=&4\tbeta^{\phi(1)},\cr}}
where 
\eqn\Ee{
\tbeta^{\phi}=2\bbeta^{\phi}-\frak{1}{2}g^{\m\n}\bbeta^g_{\m\n}.}
To consider $O(d,d)$ duality, we specialise to a metric with signature
$(-,+,+,\ldots,+)$ and we consider a $\s$-model where the fields depend only
on the first co-ordinate $t$. We can then bring $g$ and $b$ to the 
block-diagonal form
\eqn\Ef{
g_{\m\n}=\pmatrix{g_{00}&0\cr0&G(t)\cr},\qquad b_{\m\n}=\pmatrix{0&0\cr0&B(t)}.}
(For the discussion of $O(d,d)$ invariance at one loop, one can take 
$g_{00}=-1$; but at higher loops we need to consider a general $g_{00}$ at
intermediate stages, returning to $g_{00}=-1$ at the end of the computation. In
fact, we shall only retain a general $g_{00}$ at points where it will leave an
imprint even after setting $g_{00}=-1$--for instance, where it is acted upon
by a $\partial\over{\partial g_{00}}$.)
It was shown in Ref.~\mva\ that the one-loop action may then be written as 
\eqn\Eg{
\Gamma^{(1)}=-\int dte^{-\Phi}g_{00}^{-\frak12}\left(\Phidot^2
+\frak18\Tr[\Mdot\eta\Mdot\eta]\right),}
where 
\eqn\Eh{
\Phi=2\phi-\frak12\ln\det G,}
$\eta$ is the metric for the $O(d,d)$ group in non-diagonal form given by
\eqn\Ei{
\eta=\pmatrix{0_d&1_d\cr 1_d&0_d\cr},}
and
\eqn\Ej{
M=\pmatrix{G^{-1}&-G^{-1}B\cr 
BG^{-1}&G-BG^{-1}B\cr}.}
The $O(d,d)$ group is represented by matrices $\Omega$ such that
\eqn\Ek{
\Omega\eta\Omega^T=\eta;}
the action in Eq.~\Eg\ is invariant under the action of the $O(d,d)$ group
with 
\eqn\El{
M\rightarrow M'=\Omega^TM\Omega,\qquad \Phi\rightarrow\Phi, \qquad g_{00}
\rightarrow g_{00}.}
The matrix $M$ has two important properties: $M$ is symmetric, and 
$M\in O(d,d)$. 

As we mentioned earlier, the 
invariance of the action does not 
manifestly guarantee that the conformal invariance conditions
transform appropriately.
In the $O(d,d)$ case, even at one loop, where the conformal invariance 
conditions are simply
related 
to the action by Eq.~\Ed, it does not seem immediately obvious how the correct 
properties for the $\bbeta$ follow from the invariance of the action;
and at higher loops, where the relation of the $\bbeta$-functions to the action
is more complicated, it is still less clear. Therefore in this paper we
shall explicitly check the transformation properties of the $\bbeta$-functions. 

As in the case of 
the action, the $\beta$-functions are most conveniently discussed in
terms of the matrix $M$. Defining $\beta^M=\m{d\over{d\m}}M$, we have
from Eq.~\Ej\
\eqn\Eo{
\beta^M=\pmatrix{-G^{-1}\beta^G G^{-1}&G^{-1}(\beta^GG^{-1}B-\beta^B)\cr
(\beta^B-G^{-1}\beta^G)G^{-1}&\beta^G-\beta^BG^{-1}B+BG^{-1}\beta^GG^{-1}B
-BG^{-1}\beta^G\cr}.}
We then define 
\eqn\Ep{\eqalign{
\bbeta^M=&\beta^M-\frak12(\dot\phi+S) \Mdot\cr
=&\pmatrix{-G^{-1}\bbeta^GG^{-1}&G^{-1}(\bbeta^GG^{-1}B-\bbeta^B)\cr
(\bbeta^B-G^{-1}\bbeta^G)G^{-1}&\bbeta^G-\bbeta^BG^{-1}B+BG^{-1}\bbeta^GG^{-1}B
-BG^{-1}\bbeta^G\cr},\cr}}
where we assume (as will always be the case in the present calculation)
that $S_i=0$ and $S_0\equiv S$.

We start with the one-loop $\bbeta$-functions. As mentioned before, these have
already been demonstrated to have the correct properties, 
but we shall repeat the 
check in order to demonstrate our formalism in operation.
Upon specialising the general results in Eqs.~\Ea, \Eb\ to the forms for 
$g_{\m\n}$ and $b_{\m\n}$ given in Eq.~\Ef, we find that
the $\bbeta$-functions for $G$ and $B$ are given at one loop by 
\eqn\Em{\eqalign{
\bbeta^{G(1)}
=&\frak12(-g^{00})G(X-W^2-\tW^2-\Phidot W-\frak12g^{00}\gdot_{00}W),\cr
\bbeta^{B(1)}=&\frak12(-g^{00})G(\tX-W\tW-\tW W-\Phidot \tW
-\frak12g^{00}\gdot_{00}\tW),\cr}}
where 
\eqn\En{
W=G^{-1}\Gdot,\qquad \tW=G^{-1}\Bdot,\qquad X=G^{-1}\Gdotdot,\qquad
\tX=G^{-1}\Bdotdot.}
As mentioned earlier, to demonstrate the invariance at two and three loops. 
we need to retain a general $g_{00}$ for the moment, though we shall set 
$g_{00}=-1$ at the end of the calculation. 
Substituting Eq.~\Em\ in Eq.~\Ep, we find that we can write 
\eqn\Eq{
\bbeta^{M(1)}=\frak12(\Mddot+\Mdot\eta\Mdot\eta M-\Phidot\Mdot).}
(Expressions for $\Mdot$ and $\Mddot$ are given in the Appendix.) 
The $O(d,d)$ invariance for $\bbeta^G_{ij}$ and $\bbeta^B_{ij}$ is then 
manifest,
for clearly when $M$ and $\Phi$ transform according to Eq.~\El, $\bbeta^{M(1)}$
transforms according to 
\eqn\Er{
\bbeta^{M}\rightarrow\bbeta^{M}(M')=\Omega^T\bbeta^{M}(M)\Omega .}
Now if Eq.~\Er\ is satisfied, then since from Eqs.~\El, \Ep, we also have 
$\bbeta^{M'}(M')=\Omega^T\bbeta^{M}(M)\Omega$, we deduce that 
$\bbeta^{M}(M)=\bbeta^{M'}(M)$, i.e. $\bbeta^M$ is 
form-invariant under the $O(d,d)$ transformation. A 
solution of the conformal invariance conditions with $\bbeta^M(M)=0$
automatically satisfies $\bbeta^{M'}(M')=0$; but now we also have 
$\bbeta^M(M')=0$, i.e. the transformed $M'$ is a solution of the {\bf same} 
conformal invariance conditions as $M$. Eq.~\Er\ will be our touchstone for 
$O(d,d)$ invariance at two and three loops as well. 
 
We also find
\eqn\Es{
\bbeta^{g(1)}_{00}=\Phiddot+\frak18\Tr[\Mdot\eta\Mdot\eta]-
\frak12g^{00}\gdot_{00}\Phidot}
and 
\eqn\Et{
\bbeta^{\Phi(1)}=2\bbeta^{\phi(1)}-\frak12G^{ij}\bbeta^{G(1)}_{ij}=(-g^{00})
(\frak12\Phiddot-\frak12\Phidot^2-\frak14g^{00}\gdot_{00}\Phidot).}
We note here that the trace of an even number of products of $M\eta$ and its
derivatives is manifestly invariant under Eq.~\El\ while the trace of an odd 
number of such products is zero, and therefore $\bbeta^{g(1)}_{00}$ and 
$\bbeta^{\Phi(1)}$ are $O(d,d)$-invariant. 
\newsec{The two-loop case}
In this section we shall show that Eq.~\Er\ continues to hold,
and that $\bbeta^{g(1)}_{00}$ and 
$\bbeta^{\Phi(1)}$ are $O(d,d)$-invariant, at two loops. At this order, however,
we find that to make the invariance manifest we need to make field
redefinitions, as has already been found in the case of the two-loop action in 
Ref.~\meis. 
 
The two-loop $\beta$-functions are given by\ref\tors{R.R.~Metsaev and 
A.A.~Tseytlin, \plb191 (1987) 115; \npb293 (1987) 385\semi
D.~Zanon, \plb191 (1987) 363\semi
C.M.~Hull and P.K.~Townsend, \plb191 (1987) 115\semi
I.~Jack and D.R.T.~Jones, \plb200 (1988) 453}
\eqn\Eu{\eqalign{
\beta^{g(2)}_{\m\n}
=&\frak12R_{\m\k\r\s}R_{\n}{}^{\k\r\s}+\frak{1}{24}\del_{\m}H_{\k
\r\s}\del_{\n}H^{\k\r\s}+\frak18\left(1-2\l\right)\del_{\k}H_{\m\r\s}
\del^{\k}H_{\n}{}^{\r\s}\cr
&+\frak14\l R^{\k}{}_{\m\n\l}H_{\k\r\s}H^{\l\r\s}
+\frak12(3-2\l)R^{\k}{}_{\l\r\m}H_{\n\s\k}H^{\l\r\s}\cr
&+\frak12(\l-1)R_{\k\l\r\s}
H_{\m}{}^{\k\r}H_{\n}{}^{\l\s}\cr
&+\frak{1}{16}H_{\k\l\r}H_{\s}{}^{\l\r}H^{\k}{}_{\m\t}H_{\n}{}^{\t\s}
+\frak{1}{16}H_{\m\k\l}H_{\n\r\s}H_{\t}{}^{\k\r}H^{\t\l\n},\cr
\beta^{b(2)}_{\m\n}=&-R_{\k\l\r\m}\del^{\l}H_{\n}{}^{\k\r}+\frak18\del_{\k}\left(
H_{\m\l\r}H_{\n}{}^{\l\s}\right)H_{\s}{}^{\k\r}\cr
&-\frak14(\l-1)\del_{\k}\left(H_{\s}{}^{\l\r}H_{\l\r\m}\right)H_{\n}{}^{\s\k}
+\frak18\l\del_{\k}H_{\m\n\s}H^{\k}{}_{\l\r}H^{\s\l\r},\cr
\beta^{\phi(2)}=&-\frak18R_{\k\l\r\s}R^{\k\l\r\s}-\frak{1}{24}
\left(1+6\m\right)
\del_{\k}H_{\l\r\s}\del^{\k}H^{\l\r\s}\cr
&+\frak18(5+12\m)R_{\k\l\r\s}H_{\t}{}^{\l\s}H^{\t\k\r}
-\frak{1}{16}\left(\frak14+\l+3\m\right)H_{\k\r\s}H_{\l}{}^{\r\s}H^{\k\t\o}
H^{\l}{}_{\t\o}\cr
&-\frak{5}{192}H_{\k\l\r}H^{\k}{}_{\s\t}H_{\o}{}^{\l\s}H^{\o\r\t}
\cr&+\frak{1}{8}\l\del_{\k}\del_{\l}\phi H^{\k\r\s}H^{\l}{}_{\r\s},\cr}}
with
\eqn\sdef{
 S_{\m}=-\frak{1}{24}(\l+3\m)\del_{\m}(H_{\r\s\t}H^{\r\s\t}).} 
Here, $\l$ and $\m$ represent the effects of field redefinitions of 
the form
\eqn\Ev{
\delta g_{\m\n}=\l H_{\m\r\s}H_{\n}{}^{\r\s},\qquad 
\delta \phi=-\frak12\m H_{\r\s\t}H^{\r\s\t}.}
With $g_{\m\n}$ and $b_{\m\n}$ as given by Eq.~\Ef,
we find, using identities given in the Appendix, 
\eqn\Ew{\eqalign{
\bbeta^{G(2)}=&\frak{1}{32}\Bigl\{-4XW^2+4X^2+4\l W\tW^2W
+8(1-\l)W^2\tW^2-8(1-\l)W\tW W\tW\cr
&+4(1-2\l)\tX(W\tW+\tW W)-8(1-2\l)\tX^2-4(3-2\l)X\tW^2
\cr&-8(1-\l)\tW X\tW+4(1-\l)\tW W^2\tW+4\tW^4\cr
&-(1-2\lambda)(\tr[\tW^2W^2+\tr[W^2]\tW^2)+\tr[W^2]W^2+\tr[\tW^2]\tW^2\cr
&-2\l\tr[\tW^2]X-2\l\tr[\tW^2W]W+2(\l+3\m)\pa_0\tr(\tW^2)W\Bigr\}
+\hbox{transpose},   \cr}}
\eqn\Ewa{\eqalign{
\bbeta^{B(2)}=&\frak18\Bigl\{2X\tX-W^2\tX-X(W\tW+\tW W)
+2\l W\tW^3\cr&+(1-2\l)\tX\tW^2+2(1-\l)\tW\tX\tW-2(1-\l)\tW W\tW^2\cr
&+\frak12\tr[W^2]W\tW-\frak12(1-2\l)\tr[\tW^2]W\tW-\frak12\lambda\tr[\tW^2]\tX
\cr
&-\frak12\lambda\tr[W\tW^2]\tW+\frak12(\l+3\m)\pa_0\tr(\tW^2)\tW
\Bigr\}-\hbox{transpose},    \cr}}
\eqn\Ewb{\eqalign{
\bbeta^{g(2)}_{00}=&\frak14\Tr\bigl[-X^2+XW^2-\frak14W^4
+(1-\l)\tX^2-(1-\l)(\tX W\tW+\tX \tW W)\cr
&+(1-\l)W\tW W\tW-W^2\tW^2+(3-\l)X\tW^2-\frak34\tW^4\cr
&-(\l+3\m)\pa_0^2(\tW^2)\bigr],   \cr}}
\eqn\Ewc{\eqalign{
\bbeta^{\Phi(2)}=&\frak{1}{128}\Tr[\Mdot\eta\Mdot\eta\Mdot\eta\Mdot\eta]
-\frak{1}{256}\left(\Tr[\Mdot\eta\Mdot\eta]\right)^2\cr
&-\frak14(\l+3\m)\Tr[\tX^2+2X\tW^2+W\tW W\tW-\tX(W\tW+\tW W)\cr
&-W^2\tW^2
-\tW^4+\frak14\tr[\tW^2]\left(\tr[W^2]-\tr[\tW^2]\right)
-\tr[\tX\tW-\tW^2W]\Phidot]
\cr&-\frak18\l\left(\tr[\tW^2]\Phiddot+\tr[\tW^2W]\Phidot\right).\cr }}
The $O(d,d)$ invariance is not immediately manifest at two loops; for
instance, $\bbeta^{g(2)}_{00}$ and $\bbeta^{\Phi(2)}$ cannot be written in 
terms of the traces of products
of an even number of $M\eta$ and its derivatives. However, we may take 
advantage of the possibility of redefining the fields by
\eqn\redef{\eqalign{
(G_r)_{ij}=&G_{ij}+\d G_{ij}, \quad (B_r)_{ij}=B_{ij}+\d B_{ij}
\quad (g_r)_{00}=g_{00}+\d g_{00}, \cr
\Phi_r=&\Phi+\d \Phi,\quad S_r=S+\d S ,\cr}}
with $M_r$, $\beta^{M_r}$ and $\bbeta^{M_r}$ defined as in Eqs.~\Ej, \Eo\ and 
\Ep, but with $G$ replaced by $G_r$, etc.
Of course, in order to maintain the global $O(d,d)$ symmetry, the variations in 
Eq.~\redef\ should only depend on $t$. 
Note that here again we have included a general $g_{00}$. The idea that duality 
invariance might require corrections at higher orders was put forward in 
Ref.~\ref\arka{A.A.~Tseytlin, \mpla6 (1991) 1721}, and an early example 
in the current context for a particular string background was given in Ref.~\ref
\jal{J.~Panvel, \plb284 (1992) 50}. The two-loop corrections required for 
$T$-duality of the general string effective action were obtained
in Ref.~\ref\kal{N.~Kaloper and K.A.~Meissner, \prd56 (1997) 7940}, and the 
invariance of the $\beta$-functions was discussed in the torsion-free case in 
Refs.~\haag, \ref\schi{
P.E.~Haagensen, K.~Olsen, R.~Schiappa,
\prl79 (1997) 3573}.

The changes in Eq.~\redef\ induce 
corresponding modifications in the $\bbeta$-functions according to 
\eqn\dbeta{\eqalign{
\delta\bbeta^G
=&\left(\beta^G_{kl}{\pa\over{\pa G_{kl}}}
+\beta^g_{00}{\pa\over{\pa g_{00}}}+
\beta^B_{kl}{\pa\over{\pa B_{kl}}}\right)\delta G\cr
&-\Delta\beta^G+\left(\delta S
+\frak12\delta\Phidot\right)W\cr
\delta\bbeta^B=&\left(\beta^G_{kl}{\pa\over{\pa G_{kl}}}
+\beta^g_{00}{\pa\over{\pa g_{00}}}+
\beta^B_{kl}{\pa\over{\pa B_{kl}}}\right)\delta B\cr
&-\Delta\beta^B+(\delta S+\frak12\delta\Phidot)\tW\cr
\delta\bbeta^g_{00}=&\left(\beta^G_{kl}{\pa\over{\pa G_{kl}}}+
\beta^B_{kl}{\pa\over{\pa B_{kl}}}\right)\delta g_{00}\cr
&-\Delta\beta^g_{00}-2\left(\delta{\dot S}+\frak12\delta\Phidot\right)\cr
\delta\bbeta^{\Phi}=&\left(\beta^G_{kl}{\pa\over{\pa G_{kl}}}+
\beta^B_{kl}{\pa\over{\pa B_{kl}}}+\beta^g_{00}{\pa\over{\pa g_{00}}}\right)
\delta \Phi-\frak12\delta\Phiddot-\frak14\tr[W]\delta\Phidot\cr
&-\frak12\delta g^{00}[\Phiddot+\frak12\tr[W]\Phidot]
-\frak14\delta\gdot_{00}\Phidot+\Phidot\left(\delta S
+\frak12\delta\Phidot\right),\cr}}
where 
\eqn\ddbeta{
\Delta\beta^G=\beta^{G_r}(G_r,B_r,(g_r)_{00})-\beta^{G_r}(G,B,g_{00}),}
and similarly for $\Delta\beta^B$, $\Delta\beta^g_{00}$. 
Note that we could restore 
$g_{00}=-1$ by making a co-ordinate redefinition; such 
a diffeomorphism leaves the $\bbeta$-functions unaltered\ark\ref\hugh{H.~Osborn,
\npb294 (1987) 595}, 
and the overall changes in the $\bbeta$-functions would therefore remain
as given by Eq.~\dbeta. We should also mention at this point that we could
alternatively (and equivalently) 
keep the fields $G_{ij}$, $B_{ij}$, $\Phi$ and $S$ fixed, and 
instead change the duality transformations Eq.~\El. However, the approach we 
have adopted is more convenient calculationally.
  
Taking in Eq.~\dbeta
\eqn\Ex{\eqalign{
\delta G=&\frak14(-g^{00})G[W^2-(1-2\lambda)\tW^2],\qquad 
\delta g_{00}=-\frak14\l\tr[\tW^2],\cr 
\delta B=&\frak14(-g^{00})G(W\tW+\tW W),\qquad 
\delta \Phi=-\frak14(\l+3\m)(-g^{00})\tr[\tW^2],\cr
\delta S=&\frak{1}{16}\lambda(-g^{00})(2\tr[\tX\tW-2\tW^2W]+\tr[\tW^2]\tr[W])
,\cr}}
we find
\eqn\dbetab{\eqalign{
\d\bbeta^{G(2)}=&\frak18\Bigl(-X^2
+(1-2\lambda)\tX^2+2XW^2+2\lambda\tX W\tW\cr
&+(1-\lambda)[2\tW X\tW+2X\tW^2-2\tX\tW W-2W^2\tW^2+2W\tW W\tW\cr
&+W\tW^2 W]+(\lambda-2)\tW W^2\tW-W^4\Bigr)\cr
&+\frak{1}{32}\Bigl((1-2\lambda)(\tr[\tW^2]W^2+\tr[W^2]\tW^2)
-\tr[W^2]W^2-\tr[\tW^2]\tW^2\cr
&+2\l\tr[\tW^2]X+2\l\tr[\tW^2W]W\Bigr)+\hbox{transpose},\cr}}
\eqn\dbetac{\eqalign{
\d\bbeta^{B(2)}=&\frak14\Bigl(-X\tX+\tX W^2+XW\tW+(1-\lambda)[-\tX\tW^2\cr
&+\tW^3 W+\tW^2 W\tW-\tW\tX\tW]-W^3\tW\Bigr)\cr
&+\frak18\Bigl[-\frak12\tr[W^2]W\tW
+\frak12(1-2\l)\tr[\tW^2]W\tW+\frak12\lambda\tr[\tW^2]\tX\cr
&+\frak12\lambda\tr[W\tW^2]\tW\Bigr]-\hbox{transpose},\cr}}
\eqn\dbetad{\eqalign{
\d\bbeta^{g(2)}_{00}=&\frak18\tr[2XW^2-2W^4-2(1-\lambda)X\tW^2-2(1+\lambda)
\tX(\tW W+W\tW)\cr
&+4W^2\tW^2+(1+2\lambda)W\tW W\tW-\tW^4+2\lambda\tX^2].\cr}}
\eqn\dbetae{\eqalign{
\d\bbeta^{\Phi(2)}=&-\frak14(\l+3\m)
\Bigl(\tr[-2X\tW^2+W^2\tW^2+\tW^4+\tX(W\tW+\tW W)\cr
&-W\tW W\tW-\tX^2]
+\Phidot\tr[\tX\tW-\tW^2W]+\frak14\tr[\tW^2-W^2]\tr[\tW^2]\Bigr)\cr
&+\frak18\lambda\left(\tr[\tW^2]\Phiddot+\tr[\tW^2W]\Phidot\right).\cr}}

Combining Eqs.~\Ew--\Ewc\ with Eqs.~\dbetab--\dbetae, substituting in Eq.~\Ep,
and using the results in the Appendix, we now find that we have up to this order
\eqn\Eta{\eqalign{
\bbeta^{M_r}=&\frak12(\Mrddot+\Mrdot\eta\Mrdot\eta M_r-\Phidot\Mrdot)\cr
&-\frak18(\Mrddot\eta\Mrdot\eta\Mrdot+\Mrdot\eta\Mrdot\eta\Mrddot)
-\frak14\Mrdot\eta\Mrdot\eta\Mrdot\eta\Mrdot\eta M_r,\cr
\bbeta^{g_r}_{00}=&\left(\Phiddot+\frak18\Tr[\Mrdot\eta\Mrdot\eta]-
\frak12g^{00}\gdot_{00}\Phidot\right)\cr
&+\frak18\Tr[\Mrddot\eta\Mrddot\eta]-\frak{5}{32}\Tr[\Mrdot\eta
\Mrdot\eta\Mrdot\eta\Mrdot\eta],\cr
\bbeta^{\Phi_r}=&(-g^{00})
(\frak12\Phiddot-\frak12\Phidot^2-\frak14g^{00}\gdot_{00}\Phidot)\cr
&+\frak{1}{128}\Tr[\Mrdot\eta\Mrdot\eta\Mrdot\eta\Mrdot\eta]
-\frak{1}{256}\left(\Tr[\Mrdot\eta\Mrdot\eta]\right)^2.\cr}}
The $O(d,d)$ invariance is now manifest; as at one loop, when $M_r$ and $\Phi$
transform according to Eq.~\El, $\bbeta^{M_r}$ transforms according to Eq.~\Er,
and $\bbeta^{(g_r)}_{00}$ and $\bbeta^{\Phi_r}$ are invariant. 

The calculation of Ref.~\meis\ uses the action
corresponding to a scheme with $\lambda=\mu=0$; this action was singled out 
in Ref.~\ref\ghost{I.~Jack and D.R.T.~Jones, \npb303 (1988) 260} as being the 
unique ghost-free two-loop string effective action in the presence of torsion,
and in the present context appears to lead to the most economical 
demonstration of invariance, although its use is not mandatory. 
In this scheme, although there
is no need to redefine $g_{00}$, it is certainly still necessary to include a
general $g_{00}$ at intermediate stages of the calculation for the 
$\bbeta$-functions (due to the ${\pa\over{\pa g_{00}}}$ term in Eqs.~\dbeta), 
whereas the 
invariance of the action can be shown with $g_{00}=-1$ throughout. This is a
confirmation that the invariance of the $\bbeta$-functions is not
simply a consequence of the invariance of the action.
Note that the parts of the redefinitions in Eq.~\Ex\ involving $\lambda$ and 
$\mu$ 
are essentially undoing those in Eq.~\Ev. Nevertheless, it is a valuable 
exercise to perform
the calculation for general $\lambda$ and $\mu$ because it gives a
foretaste of the kinds of field redefinition we shall be obliged to use in the
three-loop case.
  
\newsec{The three-loop case}
In this section, we shall show the $O(d,d)$ invariance of the 
$\bbeta$-functions for fields only depending on $t$ at three
loops in the absence of torsion.
At three loops, the $\beta$-functions for a general theory 
are given by\ref\nour{S.J.~Graham, \plb197 (1987) 543\semi
A.P.~Foakes and N.~Mohammedi, \plb198 (1987)359; \npb306 (1988) 343} 
\eqn\Eu{\eqalign{
\beta^{g(3)}_{\m\n}=&\frak18\del_{\r}R_{\m\s\k\l}
\del^{\r}R_{\n}{}^{\s\k\l}-\frak{1}{16}\del_{\m}R_{\r\s
\k\l}\del_{\n}R^{\r\s\k\l}-\frak12
R_{\m\r\s\tau}R_{\n\k\l}{}^{\tau}R^{\r\l\s\k}\cr
&-\frak38R_{\m\r\s\n}R^{\r\k\l\tau}R^{\s}{}_{\k
\l\tau},\cr}}
and\ref\jjr{I.~Jack, D.R.T.~Jones and D.A.~Ross, \npb307 (1988) 531}
\eqn\Eua{
\beta^{\Phi(3)}
=-\frak{3}{16}R^{\m\l\r\s}R^{\n}{}_{\l\r\s}\del_{\m}\del_{\n}\phi
+\frak{1}{32}R_{\m\n\r\s}R^{\r\s\k\l}R_{\k\l}{}^{\m\n}
-\frak{1}{24}R_{\m\n\r\s}R^{\k\s\l\n}R_{\k}{}^{\r}{}_{\l}{}^{\m},}
with\jjr
\eqn\Eub{
 S^{(3)}_{\m}=\frak{1}{64}\del_{\m}\left(R_{\k\l\r\s}R^{\k\l\r\s}\right).}
Specialising to $g_{\m\n}$ as in Eq.~\Ef, we find, with the help of various 
identities given in the Appendix,  
\eqn\Eub{\eqalign{
\bbeta^{G(3)}=&\frak{1}{32}\Bigl\{-Y^2+X^3-2WX^2W-XW^2X+2YXW+2YWX-2YW^3\cr
&-WXWX+2WXW^3\cr
&+(\tr[XW^2]-\frak34\tr[X^2]-\frak38\tr[W^4]-\frak18(\tr[W^2])^2)W^2\cr
&+\frak32\tr[X^2-XW^2+\frak14W^4]X+\tr[W^2](XW^2-\frak14W^4-\frak34X^2)\cr
&+\frak34(\tr[XW-W^3])(W^3-2XW)\cr
&+\frak18\Bigl(-\tr[XW]\tr[W^2]+\tr[W^5]
-4\tr[XY]+2\tr[YW^2]+\frak52\tr[W^2]\tr[W^3]\cr
&+14\tr[X^2W]-13\tr[XW^3]\Bigr)W\Bigr\}+\hbox{transpose}.\cr}}
\eqn\Euc{\eqalign{
\bbeta^{G(3)}_{00}
=&\frak{1}{64}\tr[-18X^3+4XZ-2ZW^2+4Y^2+13YW^3-14Y(XW+WX)\cr
&+46X^2W^2+13XWXW-39XW^4+6W^6]\cr
&+\frak{1}{256}\tr[W^2]\tr[4YW-22XW^2+4X^2+13W^4]
\cr&+\frak{1}{64}\tr[XW-W^3]\tr[3XW-2W^3].\cr}}
\eqn\Eud{\eqalign{
\bbeta^{\Phi(3)}=&\frak{1}{64}\tr[2Y^2+2X^3-4Y(XW+WX)+4YW^3+3XWXW-2XW^4
\cr
&-\frak23W^6]
+\frak{1}{128}\tr[2YW^2-4XY+14X^2W-13XW^3+W^5]\Phidot\cr
&+\frak{3}{128}\tr[4X^2-4XW^2+W^4]\Phiddot
-\frak{1}{256}\tr[4XW^2-3W^4]\tr[W^2]\cr
&+\frak{1}{128}\left(3\tr[W^3]^2-7\tr[XW]\tr[W^3]+4\tr[XW]^2\right)
\cr
&-\frak{1}{256}\tr[2XW-5W^3]\tr[W^2]\Phidot.\cr}}
Once again, the $O(d,d)$-invariance is far from manifest, and we are obliged
to resort to field redefinitions.
Using redefinitions as in Eqs.~(A.11)--(A.14), we find variations of 
the $\bbeta$-functions as in Eqs.~(A.15)--(A.18). Taking the particular values 
of the coefficients as
given in Eq.~(A.19), we find, on combining Eqs.~(A.15)--(A.18) with 

Eqs.~\Eub--\Eud, and 
substituting in Eq.~\Ep, that we can write
\eqn\Ey{\eqalign{
\bbeta^{M_r(3)}
=&\frak{1}{32}\Bigl(\Mrt\eta\Mrdot\eta\Mrdot\eta\Mrdot\eta M_r
+\Mrdot\eta\Mrdot\eta\Mrdot\eta\Mrt\eta M_r
-\Mrddot\eta\Mrddot\eta\Mrddot\Bigr)\cr
&-\frak{1}{16}\Mrt\eta\Mrddot\eta\Mrdot
-\frak{1}{32}\left(\Mrddot\eta\Mrddot\eta\Mrdot\eta\Mrdot\eta M_r+
\Mrdot\eta\Mrdot\eta\Mrddot\eta\Mrddot\eta M_r\right)\cr
&-\frak{3}{64}\left(\Mrddot\eta\Mrdot\eta\Mrddot\eta\Mrdot\eta M_r
+\Mrdot\eta\Mrddot\eta\Mrdot\eta\Mrddot\eta M_r\right)\cr
&+\frak{1}{32}\Mrddot\eta\Mrdot\eta\Mrdot\eta\Mrddot\eta M_r
-\frak{3}{32}\Mrdot\eta\Mrddot\eta\Mrddot\eta\Mrdot\eta M_r\cr
&+\frak{3}{16}\left(\Mrddot\eta\Mrdot\eta\Mrdot\eta\Mrdot\eta\Mrdot
+\Mrdot\eta\Mrdot\eta\Mrdot\eta\Mrdot\eta\Mrdot\eta
\Mrdot\eta M_r\right)\cr
&+\frak{1}{128}\tr[\Mrdot\eta\Mrdot\eta]\left(\Mrddot\eta\Mrdot\eta\Mrdot
+\Mrdot\eta\Mrdot\eta\Mrdot\eta\Mrdot\eta M_r\right)\cr
&+\frak{1}{32}\Bigl(\tr[\Mrt\eta\Mrdot\eta
+\Mrdot\eta\Mrdot\eta\Mrdot\eta\Mrdot\eta]
\cr 
&+\frak14\tr[\Mrdot\eta\Mrdot\eta]^2\Bigr)
[\Mrddot+\Mrdot\eta\Mrdot\eta M_r]\cr
&-\frak{1}{16}\tr[\Mrddot\eta\Mrdot\eta]\left(\Mrt+3\Mrddot\eta\Mrdot\eta M_r
+\frak{33}{16}\Mrdot\eta\Mrdot\eta\Mrdot\right)\cr
&+\frak{1}{256}\Bigl(\tr[16\Mrt\eta\Mrddot\eta-61\Mrddot\eta\Mrdot\eta\Mrdot\eta
\Mrdot\eta]\cr
&-4\tr[\Mrdot\eta\Mrdot\eta]\tr[\Mrddot\eta\Mrdot\eta]\Bigr)\Mrdot\cr
&+\hbox{transpose},}}
\eqn\Eza{\eqalign{
\beta^{g(3)}_{00}
=&-\frak{33}{128}\tr[\Mrddot\eta\Mrdot\eta\Mrddot\eta\Mrdot\eta]
+\frak{59}{128}\tr[M_r^{(3)}\eta\Mrdot\eta\Mrdot\eta\Mrdot\eta]\cr
&+\frak{3}{32}\tr[M_r^{(3)}\eta M_r^{(3)}\eta]
+\frak{15}{128}\tr[(\Mrdot\eta)^6]\cr
&-\frak{1}{256}\tr[\Mrdot\eta\Mrdot\eta]\tr[\Mrddot\eta\Mrddot\eta]
-\frak{9}{256}\left(\tr[\Mrddot\eta\Mrdot\eta]\right)^2,\cr}}
\eqn\Ezb{\eqalign{
\beta^{\Phi(3)}
=&-\frak{29}{256}\tr[\Mrddot\eta\Mrdot\eta\Mrdot\eta\Mrdot\eta]
\Phidot
-\frak{1}{512}\tr[\Mrddot\eta\Mrdot\eta]\tr[\Mrdot\eta\Mrdot\eta]\Phidot\cr
&+\frak{1}{512}\tr[(\Mrdot\eta)^4]\tr[(\Mrdot\eta)^2]
-\frak{1}{384}\tr[(\Mrdot\eta)^6]\cr
&+\frak{3}{256}\left(\tr[\Mrddot\eta\Mrdot\eta]\right)^2
-\frak{1}{512}(\tr[\Mrdot\eta\Mrdot\eta])^3.\cr}}
In these equations $M_r$ is defined by Eq.~\Ej, but with $B=0$.
Once again, 
the $O(d,d)$ invariance is now manifest; when $M$ and $\Phi$
transform according to Eq.~\El, $\bbeta^{M_r}$ transforms according to Eq.~\Er,
and $\bbeta^{(g_r)}_{00}$ and $\bbeta^{\Phi_r}$ are invariant. 
We have tried to choose the field redefinitions in order to minimise the 
number of terms which appear here--clearly with only partial success. 
However, we should stress that it is very non-trivial and apparently 
miraculous that an 
$O(d,d)$-invariant form could be found at all, since there are many more 
constraints than there are free parameters.  
\newsec{Conclusions}
In this paper we have shown explicitly that the conformal invariance 
conditions are form-invariant under $O(d,d)$ transformations
up to two loops for the general case with torsion. In principle, it should be 
possible to compare our
results with the two-loop results of Ref.~\jal\ which were established for a 
specific
background. We have also demonstrated the $O(d,d)$ invariance up to three loops
in the torsion-free case. We should mention, however, that we expect 
at three loops that in the
presence of torsion, assuming 
that $O(d,d)$ invariance is still preserved,
the various $\bbeta$-functions will still adopt the form
of Eqs.~\Ey--\Ezb, but with $M_r$ now including $B$ as in Eq.~\Ej. The 
three-loop $\beta$-function in the presence of torsion has been calculated\ref
\ket{S.V.~Ketov, A.A.~Deriglazov and Y.S.~Prager, \npb332 (1990) 447},
but evidently the inclusion of torsion in the present computation would be 
prohibitively complex. Finally, 
it is interesting that we found it essential to keep $g_{00}$ 
as a variable during the calculation, even though we set $g_{00}=-1$ at the end
of the calculation.
It is not clear how this is accounted for in the argument for 
all-orders 
$O(d,d)$ invariance presented in Ref.~\sen, where the gauge $g_{00}=-1$ was 
chosen from the outset. Presumably  
one could in fact demonstrate $O(d,d)$-invariance with a general $t$-independent
$g_{00}$, without setting $g_{00}=-1$ at the end, though this would be 
somewhat tedious.    
\vskip 10pt
\line{\bf Acknowledgements\hfil}
S.P. was supported by a PPARC Graduate Studentship.
\appendix{A}{}
In this appendix we list various identities which were useful in our 
calculations. Firstly here are the results we used to express the two-loop
$\beta$-functions in terms of $W$, $\tW$ etc. in the two loop case:
\eqn\Aba{\eqalign{
R_{i\m\n\r}R_j{}^{\m\n\r}=&\frak14[2X^2-XW^2-W^2X]_{ij}+\frak18\tr[W^2]W^2_{ij}
,\cr
\del_iH_{\a\b\g}\del_jH^{\a\b\g}=&-\frak34\tr[\tW^2]W^2_{ij}+\frak32(W\tW^2
W)_{ij},\cr
\del_{\m}H_{i\n\r}\del^{\m}H_j{}^{\n\r}=&-\frak14\tr[\tW^2]W^2_{ij}
-\frak12\tr[W^2]\tW^2_{ij}+\frak12(W^2\tW^2+\tW^2W^2\cr
&-W\tW W\tW-\tW W\tW W-W\tW^2W+2\tX W\tW\cr
&+2\tX\tW W+2W\tW\tX+2\tW W\tX-4\tX^2)_{ij},\cr
R^{\m}{}_{ij\n}H_{\m\a\b}H^{\n\a\b}=&\frak14\tr[\tW^2](-2X_{ij}+W^2_{ij})
+\frak12(W\tW^2W)_{ij}-\frak12\tr[\tW^2W]W_{ij},\cr
R^{\m}{}_{\n\r i}H_{j\s\m}H^{\n\r\s}=&\frak14(-2X\tW^2+W^2\tW^2-W\tW W\tW)_{ij},
\cr
R_{\m\n\r\s}H_i{}^{\m\r}H_j{}^{\n\s}=&(\tW X\tW-\frak12\tW W^2\tW)_{ij},\cr
H_{\m\n\r}H_{\s}{}^{\n\r}H^{\m}{}_{i\t}H_j{}^{\t\s}=&2\tW^4_{ij}+\tr[\tW^2]
\tW^2_{ij},\cr
H_{i\m\n}H_{j\r\s}H_{\t}{}^{\m\r}H^{\t\n\s}=&2\tW^4_{ij}.\cr}}
\eqn\Abb{\eqalign{
R_{\a\b\g i}\del^{\b}H_j{}^{\a\g}=&\frak18(2W^2\tX+2X\tW W+2XW\tW-4X\tX\cr
&-W^2\tW W+W\tW W^2-\tr[W^2]W\tW)_{ij},\cr
\del_{\a}(H_{i\b\g}H_j{}^{\b\delta})H_{\delta}{}^{\a\g}
=&(2W\tW^3-\tW^2W\tW+\tW W\tW^2
-2\tX\tW^2)_{ij},\cr
\del_{\a}(H_{\b}{}^{\g\delta}H_{\g\delta i})H_j{}^{\b\a}
=&(2\tX\tW^2+2\tW\tX\tW-W\tW^3
-2\tW W\tW^2-\frak12\tr[\tW^2]W\tW)_{ij},\cr
\del_{\a}H_{ij\b}H^{\a}{}_{\r\s}H^{\b\r\s}=&\frak12\tr[\tW^2](W\tW+\tW W
-2\tX)_{ij}\cr
&+(W\tW^3+\tW^3 W-\tr[W\tW^2]\tW)_{ij}.\cr}}
\eqn\Abc{\eqalign{
R_{0\m\n\r}R_0{}^{\m\n\r}=&-\frak18\tr[4X^2-4XW^2+W^4],\cr
\del_0H_{\m\n\r}\del_0H^{\m\n\r}=&3\del_{\m}H_{0\n\r}\del^{\m}H_0{}^{\n\r}=
3\tr[\tX^2-\tX(W\tW+\tW W)\cr
&+\frak12(W\tW W\tW+W^2\tW^2)],\cr
R^{\m}{}_{00\n}H_{\m\r\s}H^{\n\r\s}=&2R^{\m}{}_{\n\r0}H_{0\s\m}H^{\n\r\s}=
\tr[X\tW^2-\frak12W^2\tW^2],\cr
R_{\m\n\r\s}H_0{}^{\m\r}H_0{}^{\n\s}=&-\frak14\tr[\tW W\tW W],\cr
H_{\m\n\r}H_{\s}{}^{\n\r}H^{\m}_{0\t}H_0{}^{\t\s}=&
2H_{0\m\n}H_{0\r\s}H_{\t}{}^{\m\r}H^{\t\n\s}=-2\tr[\tW^4].\cr}}
Next, here are the identities we used to write $\bbeta^M$ in terms of $M$ at
two loops:
\eqn\Aa{\eqalign{
\Mdot=&\pmatrix{-WG^{-1}&WP-\tW\cr G(\tW-PW)G^{-1}&
G(W+PWP-\tW P-P\tW)\cr}, \cr
\Mddot=&\pmatrix{AG^{-1}&-AP+B\cr
G(PA+B^T)G^{-1}&G(X-B^TP+PB+PAP)\cr},\cr}}
where
\eqn\Ab{
P=G^{-1}B, \qquad A=2W^2-X, \qquad B=2W\tW-\tX.}
\eqn\Ab{\eqalign{
\Tr[\Mdot\eta\Mdot\eta]=&\tr[W^2-\tW^2]\cr
\Tr[\Mdot\eta\Mdot\eta\Mdot\eta\Mdot\eta]
=&\Tr[W^4+\tW^4-4W^2\tW^2+2W\tW W\tW].\cr}}
Now here are the identities for writing the three-loop $\bbeta$-functions in 
terms of $W$, etc:
\eqn\Ac{\eqalign{
\del_{\r}R_{i\s\k\l}\del^{\r}R_{j}{}^{\s\k\l}=&-\frak12Y^2
+\frak12(YXW+XWY+YWX+WXY)\cr
&-\frak12(YW^3+W^3Y)-\frak14XW^2X-\frak38WX^2W
\cr
&+\frak14(W^2X^2+X^2W^2)-\frak12(XWXW+WXWX)\cr
&+\frak{5}{16}(WXW^3+W^3XW)-\frak38W^2XW^2+\frak38W^6\cr
&-\frak{7}{32}(\tr[W^4]W^2+\tr[W^2]W^4)-\frak38\tr[W^2]X^2-\frak14\tr[X^2]W^2\cr
&+\frak38\tr[XW^2]W^2+\frak14\tr[W^2](XW^2+W^2X)-\frak{7}{16}\tr[W^3]W^3\cr
&+\frak14\tr[W^3](XW+WX)+\frak38\tr[XW]W^3-\frak18\tr[XW](XW+WX),\cr
\del_{i}R_{\r\s\k\l}\del_{j}R^{\r\s\k\l}=&\frak12WX^2W-\frak14(W^3XW+WXW^3)
-\frak12W^2XW^2+\frak12W^6\cr
&-\frak18(\tr[W^2]W^4+\tr[W^4]W^2)-\frak12\tr[X^2]W^2+\frak12\tr[XW^2]W^2\cr
&-\frak14\tr[W^3]W^3+\frak12\tr[XW]W^3,\cr
R_{i\r\s\tau}R_{j\k\l}{}^{\tau}R^{\r\l\s\k}=&-\frak18X^3+\frak{1}{16}XW^2X
+\frak{1}{16}(X^2W^2+W^2X^2)\cr
&-\frak{1}{16}WXWX+XWXW)-\frak{1}{32}W^2XW^2+\frak{1}{32}(WXW^3+W^3XW)\cr
&+\frak{1}{32}W^6-\frak{1}{32}\tr[W^2]W^4-\frak{1}{64}\tr[W^4]W^2
+\frak{1}{64}(\tr[W^2])^2W^2\cr
&+\frak{1}{16}\tr[XW](WX+XW)+\frak{1}{64}\tr[W^3]W^3\cr
&-\frak{1}{16}\tr[XW]W^3-\frak{1}{32}\tr[W^3](XW+WX),\cr
R_{i\k\l j}R^{\k\r\s\tau}R^{\l}{}_{\r\s\tau}=&\frak14\left(\tr[X^2]-\tr[XW^2]
+\frak14\tr[W^4]\right)(\frak12W^2-X)+\frak{1}{32}\tr[W^2]W^4\cr
&+\frak18(\tr[XW^3]-\tr[X^2W]-\frak14\tr[W^2]\tr[W^3])W\cr
&+\frak18WX^2W-\frak{1}{16}(W^3XW+WXW^3),\cr
\del_i\del_j(R_{\m\n\r\s}R^{\m\n\r\s})=&-\frak14(4\tr[XY]-8\tr[X^2W]-2\tr[YW^2]
+7\tr[XW^3]\cr
&-\tr[W^5]+\tr[XW]\tr[W^2]-\tr[W^3]\tr[W^2])\cr}}
\eqn\Ad{\eqalign{
\del_{\r}R_{0\s\k\l}\del^{\r}R_{0}{}^{\s\k\l}=&\frak12\tr[Y^2]+\tr[XWXW]
+\frak38\tr[W^6]-\tr[YXW]-\tr[YWX]\cr
&+\tr[YW^3]+\frak78\tr[X^2W^2]-\frak74\tr[XW^4]+\frak{1}{16}\tr[W^4]\tr[W^2]\cr
&+\frak18\tr[W^2]\tr[X^2]+\frak{1}{16}(\tr[W^3])^2-\frak18\tr[W^3]\tr[XW]
\cr
&-\frak18\tr[W^2]\tr[XW^2],\cr
\del_{0}R_{\r\s\k\l}\del_{0}R^{\r\s\k\l}=&\tr[Y^2]-2\tr[YXW]-2\tr[YWX]
+2\tr[YW^3]\cr
&+2\tr[XWXW]+\frak32\tr[X^2W^2]-3\tr[XW^4]+\frak12\tr[W^6]\cr
&+\frak14\tr[X^2]\tr[W^2]-\frak12\tr[XW^2]\tr[W^2]+\frak14\tr[W^4]\tr[W^2]\cr
&+\frak14(\tr[W^3-XW])^2,\cr
R_{0\r\s\tau}R_{0\k\l}{}^{\tau}R^{\r\l\s\k}=&\frak{1}{32}\tr[4X^3-6X^2W^2
+2XWXW+XW^4]\cr
&-\frak{1}{64}\left\{(\tr[W^3])^2-4\tr[XW]\tr[W^3]+4(\tr[XW])^2\right\},
\cr 
R_{0\k\l 0}R^{\k\r\s\tau}R^{\l}{}_{\r\s\tau}=&\frak14\tr[X^3]
-\frak38\tr[X^2W^2]+\frak18\tr[XW^4]\cr
&+\frak{1}{16}\tr[W^2]\tr[XW^2]-\frak{1}{32}\tr[W^2]\tr[W^4],\cr
\del_0\del_0(R_{\m\n\r\s}R^{\m\n\r\s})=&\frak12\Bigl(4\tr[XZ]+4\tr[Y^2]
-14\tr[YWX]-14\tr[YXW]\cr
&-8\tr[X^3]+30\tr[X^2W^2]+15\tr[XWXW]-2\tr[ZW^2]\cr
&+13\tr[YW^3]-33\tr[XW^4]+5\tr[W^6]+\tr[YW]\tr[W^2]\cr
&+\tr[X^2]\tr[W^2]+2(\tr[XW])^2-5\tr[XW^2]\tr[W^2]\cr
&-4\tr[XW]\tr[W^3]+3\tr[W^4]\tr[W^2]+2(\tr[W^3])^2\Bigr).\cr}}
\vfill
\eject
The following are the identities needed to write $\bbeta^M$ in terms of $M$ at 
three loops (here $M$ is defined by Eq.~\Ej, but with $B=0$):
\eqn\Af{\eqalign{
 \Mt\eta\Mddot\eta\Mdot=&\pmatrix{(Y-3WX-3XW+6W^3)XW&0\cr
                                  0&Y(-X+2W^2)W\cr}\cr
\Mt\eta\Mdot\eta\Mdot\eta\Mdot\eta M=&\pmatrix{(Y-3WX-3XW+6W^3)W^3&0\cr
                                               0&YW^3\cr}\cr
\Mddot\eta\Mddot\eta\Mddot=&\pmatrix{X^3-2X^2W^2-2W^2X^2+4W^2XW^2&0\cr
                                     0&-X^3+2XW^2X\cr}\cr
\Mddot\eta\Mddot\eta\Mdot\eta\Mdot\eta M=&\pmatrix{X^2W^2&0\cr
                                          0&X^2W^2-2XW^4\cr}\cr
\Mddot\eta\Mdot\eta\Mddot\eta\Mdot\eta M=&\pmatrix{XWXW-2W^3XW-2XW^4+4W^6&0\cr
                                          0&XWXW\cr}\cr
\Mddot\eta\Mdot\eta\Mdot\eta\Mddot\eta M=&\pmatrix{XW^2X-2W^4X&0\cr
                                                 0&XW^2X-2XW^4\cr}\cr
\Mdot\eta\Mddot\eta\Mddot\eta\Mdot\eta M=&\pmatrix{WX^2W-2WXW^3&0\cr
                                                   0&WX^2W-2W^3XW\cr}\cr
\Mddot\eta\Mdot\eta\Mdot\eta\Mdot\eta\Mdot=&\pmatrix{-XW^4+2W^6&0\cr
                                         0&XW^4\cr}\cr
\Mdot\eta\Mdot\eta\Mdot\eta\Mdot\eta\Mdot\eta M=&\pmatrix{-W^6&0\cr
                                                           0&-W^6\cr}.\cr}}
\eqn\Ag{\eqalign{
\tr[M^{(4)}\eta\Mddot\eta]=&2\tr[-ZX+2WYX+2WXY+3X^3+W^2Z\cr
&-12W^2X^2-6XWXW+12W^4X]\cr
\tr[\Mt\eta\Mt\eta]=&2\tr[-Y^2+3WXY+3XWY-6W^3Y]\cr
\tr[\Mt\eta\Mdot\eta\Mdot\eta\Mdot\eta]=&2\tr[YW^3-3XW^4+3W^6]\cr
\tr[\Mddot\eta\Mddot\eta\Mdot\eta\Mdot\eta]=&2\tr[X^2W^2-2XW^4]\cr
\tr[\Mddot\eta\Mdot\eta\Mddot\eta\Mdot\eta]=&2\tr[XWXW-2XW^4+2W^6]\cr
\tr[\Mdot\eta\Mdot\eta\Mdot\eta\Mdot\eta\Mdot\eta\Mdot\eta]=&-2\tr[W^6]\cr
\tr[\Mt\eta\Mdot\eta]=&2\tr[-YW+3XW^2-3W^4]\cr}}
Finally, the following is the general form of field redefinition we need to 
consider in Eq.~\redef\ up to three loops: 
\eqn\Ada{\eqalign{
\delta G=&\frak14(-g^{00})W^2+\a_1X^2+\a_2(XW^2+W^2X)+\a_3WXW+\a_4W^4\cr
&+[\b_1\tr(W^3)+\b_2\tr(XW)]W+\tr(W^2)(\g_1W^2+\g_2X)\cr
&+\gdot^{00}(\l_1(XW+WX)+\l_2W^3),\cr}}
\eqn\Adaa{\eqalign{
\delta g_{00}=&(-g^{00})\left[\epsilon_1\tr(W^4)+\epsilon_2\tr(XW^2)+
\epsilon_3\tr(X^2)\right]
+\gdot_{00}\Bigl[\frak12\e_2\tr(W^3)\cr
&+\e_3\tr(XW)\Bigr],\cr}}
\eqn\Adb{
\delta \Phi=\kappa_1\tr[W^4]+\kappa_2\tr[XW^2]+\kappa_3\tr[X^2]
+\kappa_4(\tr[W^2])^2+\gdot_{00}(\mu_1\tr[XW]+\mu_2\tr[W^3]),}
\eqn\Adc{\eqalign{
(\delta S)_0=&\delta_1\tr[W^5]+\delta_2\tr[XW^3]+\delta_3\tr[X^2W]
+\delta_4\tr[XY]\cr
&+\delta_5\tr[YW^2]+\delta_6\tr[W^3]\tr[W^2]+\delta_7\tr[XW]\tr[W^2]\cr
&+\frak14\left[\e_1\tr(W^4)+\e_2\tr(XW^2)+\e_3\tr(X^2)\right]\tr(W).\cr}}
\vfill
\eject
These redefinitions lead using Eq.~\dbeta\ to changes in the $\bbeta$-functions
given by
\eqn\Ae{\eqalign{
\delta\bbeta^{G(3)}=&
\Bigl\{-\frak12\alpha_1Y^2+(\frak12\alpha_1-\alpha_2+\frak{1}{16})YXW\cr
&+(\frak12\alpha_1-\alpha_2)YWX\cr
&+(\frak{1}{16}-\alpha_3)WYX+(2\alpha_2-\frak{1}{16})YW^3
+(\alpha_3-\frak{1}{16})WYW^2\cr
&-(\alpha_1+\alpha_2+\frak12\alpha_3)X^3+(\alpha_1+\frak32
\alpha_3-\alpha_4-\frak18)X^2W^2\cr
&+(\alpha_2+\frak12\alpha_3-\alpha_4-\frak{5}{32})XWXW+\frak14(3\alpha_1+
6\alpha_2-2\alpha_4-\frak{1}{16})XW^2X\cr
&-\frak14(\alpha_1-2\alpha_2+2\alpha_3+2\alpha_4+\frak{3}{16})WX^2W
-(\alpha_1+\alpha_2-3\alpha_4-\frak14)XW^4\cr
&-\frak12(2\alpha_2-\alpha_3-4\alpha_4-\frak38)WXW^3+(\alpha_2-\alpha_3
+\alpha_4+\frak18)W^2XW^2\cr
&-\frak12(2\alpha_2+\alpha_3+6\alpha_4+\frak{7}{16})W^6
-\frak14\tr[W^2](\alpha_1X^2+(2\alpha_2-\frak{1}{16})XW^2\cr
&+(\alpha_3-\frak{1}{16})WXW+(\alpha_4+\frak{3}{32})W^4\Bigr\}
+\hbox{transpose}\cr
&+\Bigl[3\b_1\tr(-X^2W+2XW^3-W^5)\cr
&+\b_2\tr(-XY+YW^2-W^5+XW^3)\Bigr]W\cr
&+\Bigl[(\delta'_1+\e_1)\tr(W^5)+(\delta'_2-\e_1+\frak34\e_2)\tr(XW^3)\cr
&+(\delta'_3-\frak12\e_2+\frak12\e_3)\tr(X^2W)
+(\delta'_4-\frak12\e_3)\tr(XY)\cr
&+(\delta'_5-\frak14\e_2)\tr(YW^2)+\delta'_6\tr(W^3)\tr(W^2)
+\delta'_7\tr(XW)\tr(W^2)\Bigr]W\cr
&+\left[3\b_1\tr(XW^2-W^4)+\b_2\tr(YW+X^2-2XW^2)\right](W^2-X)\cr
&+\frak12\left[\e_1\tr(W^4)+\e_2\tr[XW^2]+\e_3\tr(X^2)\right]
(W^2-X)\cr
&-\frak12\left[\b_1\tr(W^3)+\b_2\tr(XW)\right]\tr(W^2)W\cr
&+\tr(-X^2+2XW^2-W^4)(\g_1W^2+\g_2X)\cr
&-\frak{1}{64}\tr[4X^2-4XW^2+W^4]W^2\cr
&-\tr(XW-W^3)\Bigl[2\g_1(XW+WX-2W^3)\cr
&+\g_2(2Y-WX-XW)-\frak{1}{16}W^3\Bigr]\cr
&-(\g_1+\g_2)\tr(W^2)(X^2-XW^2-W^2X+W^4)\cr
&-\frak12[\tr(W^2)]^2(\g_1W^2+\g_2X)\cr
&+\frak14\tr[Y-3XW+2W^3]\Bigl[(\alpha_1+2\lambda_1)(XW+WX)\cr
&+(2\alpha_2+\alpha_3+2\lambda_2)W^3+(\beta_2+\gamma_2)\tr[W^2]W\Bigr]\cr
&+\frak12\tr[XW-W^3][\lambda_1(XW+WX)+\lambda_2W^3],\cr}}
where
\eqn\Aebaa{\eqalign{
\delta_1'=\delta_1-2\k_1,\qquad &\delta_2'=\delta_2+2\k_1-\frak32\k_2,\cr
\d_3'=\d_3+\k_2-\k_3,\qquad &\d_4'=\d_4+\k_3,\cr
\d_5'=\d_5+\frak12\k_2, \qquad \d_6'=&\d_6-2\k_4, \qquad \d_7'=\d_7+2\k_4.\cr}}
\eqn\Aeb{\eqalign{
\delta\bbeta^{g(3)}_{00}=&\frak12\tr\Bigl\{(-\alpha_1+2\alpha_2+2\alpha_3
-8\delta_2'+8\delta_3'-8\e_2+10\e_3-\frak{35}{16})X^2W^2\cr
&+(-\alpha_1+2\alpha_2-4\delta'_2+4\delta'_3-\e_2+2\e_3-\frak{11}{16})XWXW\cr
&+(2\alpha_2+\alpha_3-4\delta'_2+12\delta'_5+4\e_1-2\e_2-\frak12)YW^3\cr
&+(\alpha_1-4\delta'_3+4\delta'_4-4\delta'_5+\e_2-2\e_3+\frak14)(YXW+XYW)\cr
&+(-6\alpha_2-3\alpha_3+4\alpha_4-20\delta'_1+16\delta'_2
-12\e_1+10\e_2-4\e_3+\frak{31}{8})XW^4\cr
&+(-4\alpha_4+20\delta'_1+8\e_1-2\e_2-\frak32)W^6+(-4\delta'_5+\e_2)ZW^2
\cr
&+(-4\delta'_4+2\e_3)ZX+(-4\delta'_3-6\e_3+\frak12)X^3-4\delta'_4Y^2\cr
&+(\beta_1+2\gamma_1-8\delta_6'-\frak12\e_2)\tr[W^3][XW-W^3]\cr
&+(3\beta_1+2\gamma_1-12\delta_6')\tr[XW^2-W^4]\tr[W^2]\cr
&+(\beta_2+2\gamma_2-8\delta_7'-\e_3)\tr[XW][XW-W^3]\cr
&+(\beta_2-4\delta_7')\tr[YW+X^2-2XW^2]\tr[W^2]+\g_2\tr[YW-XW^2]\tr[W^2]\cr
&-\frak12\tr(W^2)\Bigl[\e_1\tr(W^4)+\e_2\tr[XW^2]+\e_3\tr(X^2)\Bigr]
\Bigr\}.\cr}}
\eqn\Aec{\eqalign{
\delta\bbeta^{\Phi(3)}
=&\tr[-(\kappa_2+2\kappa_3)X^3-\kappa_3Y^2+(\kappa_3-\kappa_2)
(YXW+YWX)\cr
&+2\kappa_2YW^3+(\kappa_2-2\kappa_1)XWXW-(4\kappa_1-2\kappa_2-3\kappa_3)X^2W^2
\cr
&-(\kappa_2+2\kappa_3-12\kappa_1+\frak{1}{32})XW^4-(6\kappa_1+\kappa_2-
\frak{1}{32})W^6]\cr
&-\frak12\tr[W^2]\tr\Bigl([\kappa_1+\frak{1}{16}]W^4
+[\kappa_2-\frak{1}{16}]XW^2+\kappa_3X^2\Bigr)\cr
&-2\kappa_4\left(\tr[X^2-2XW^2+W^4]\tr[W^2]+2(\tr[XW-W^3])^2+(\tr[W^2])^3
\right)\cr
&+\Phidot\Bigl[(\d_1'+\e_1)\tr(W^5)+(\d_2'-\e_1+\frak34\e_2)\tr(XW^3)
\cr
&+(\d_3'-\frak12\e_2+\frak12\e_3)\tr(X^2W)
+(\d_4'-\frak12\e_3)\tr(XY)\cr
&+(\d_5'-\frak14\e_2)\tr(YW^2)+\d_6'\tr(W^3)\tr(W^2)\cr
&+\d_7'\tr(XW)\tr(W^2)\Bigr]
+\frak14\tr[Y-3XW+2W^3]\tr[2(\k_3-\m_1)XW\cr
&+(\k_2-2\m_2)W^3]
-\frak12\tr[XW-W^3]\tr[\m_1XW+\m_2W^3]\cr
&-\frak12\Bigl[\e_1\tr(W^4)+\e_2\tr(XW^2)+\e_3\tr(X^2)\Bigr]\Phiddot.\cr}}
In order to demonstrate $O(d,d)$ invariance, we find that we need to take
\eqn\Ezc{\eqalign{
\alpha_1=-\frak{1}{16},\qquad \alpha_2=\frak{1}{32},&\qquad \alpha_3=
\frak{1}{16}, \qquad \alpha_4=-\frak{1}{32}, \cr 
\beta_1=-\frak18, \qquad \beta_2=\frak18, &\qquad \gamma_1=
\frak{7}{64}, \qquad \gamma_2=-\frak18,\cr
\l_1=-\frak12\alpha_1,\qquad &\l_2=-(\alpha_2+\frak12\alpha_3),\cr
\delta_1=\frak{3}{16}, \qquad \delta_2=0, \qquad \delta_3=-\frak{15}{64},
\qquad \delta_4=&\frak{3}{32}, \qquad \delta_5=-\frak{3}{64}, \cr
\delta_6=-\frak{7}{256}, &\qquad \delta_7=\frak{1}{64},\cr
\epsilon_1=\frak{3}{64}, \qquad \epsilon_3=&-\epsilon_2=\frak{3}{16},\cr  
\kappa_1=\frak{1}{128},\qquad \kappa_3=-\kappa_2=\frak{1}{32},
\qquad \kappa_4=-&\frak{1}{128},
\qquad \mu_1=\kappa_3,\qquad \mu_2=\frak12\kappa_2 .\cr}}
\listrefs
\bye